\documentclass[12pt,preprint]{aastex}

\newcommand{\myemail}{wuyx@mails.thu.edu.cn}

\shorttitle{X-RAY SPECTRAL VARIABILITY IN INITIAL RISE OF
XTE~J1550-564 } \shortauthors{Wu, Liu\& Li}

\begin{document}
\title{X-RAY SPECTRAL VARIABILITY \\
IN INITIAL RISE OF XTE~J1550-564}
\author{Y. X. Wu\altaffilmark{1}, C. Z. Liu\altaffilmark{1} and T. P. Li\altaffilmark{1,~2,~3}}
\altaffiltext{1}{Department of Engineering Physics \& Center for
Astrophysics, Tsinghua University, Beijing, China; \myemail}
\altaffiltext{2}{Department of Physics \& Center for Astrophysics,
Tsinghua University, Beijing, China}
 \altaffiltext{3}{Particle Astrophysics Lab., Institute of High Energy Physics,
Chinese Academy of Sciences, Beijing}

\begin{abstract}
X-ray variability of the microquasar XTE~J1550-564 is studied with
time domain techniques for the data from the {\sl RXTE}/PCA
observation in September 8, 1998. The 2--60\,keV averaged shot is
obtained from superposing shots with one millisecond time bin
through aligning their peaks. The spectral behavior during the
averaged shot exhibits prominent differences from that observed in
Cyg~X-1. The hardness ratio of (13--60\,keV)/(2--13\,keV) or
(16--60\,keV)/(13--16\,keV) during a shot is lower or higher than
that of steady emission respectively. The correlation coefficient
between intensity and hardness ratio of (13--60\,keV)/(2--13\,keV)
or (16--60\,keV)/(13--16\,keV) is negative or positive respectively.
These results may indicate that physical processes in the low state
of XTE~J1550-564 are different from those in Cyg~X-1.
\end{abstract}

\keywords {accretion, accretion disks --- black hole physics ---
stars: individual (XTE~J1550-564) --- X-rays: stars}

\section{INTRODUCTION}
 The X-ray transient XTE~J1550-564 (Galactic longitude and latitude
 $l=325.88\,^{\circ},b=-1.83\,^{\circ}$) was discovered by the All-Sky Monitor (ASM)
 aboard the {\sl Rossi X-ray Timing Explorer (RXTE)} on September 7,
 1998 \citep{Smi98}. Shortly afterwards, its optical
 \citep{Oro98} and radio \citep{Cam98} counterparts were identified.

Observations with the Proportional Counter Array (PCA) aboard $RXTE$
were performed between September 7, 1998 and May 20, 1999, during
which XTE~J1550-564 went through multiple canonical states
\citep{Sob99,Hom01}. The light curve of 1998/1999 outburst of
XTE~J1550-564 was complex and included a slow (10 days) rise to
maximum, an intense and brief (one day) flare that occurred early in
the outburst \citep{Sob00b}, and a ''double-peaked'' profile that
roughly separated the outburst into two halves of comparable
intensity. Low frequency (0.08-18Hz) X-ray quasi-periodic
oscillations (QPOs) as well as high frequency variable (100-285Hz)
QPOs were detected during some of the {\sl RXTE} observations
\citep{Cui99,Rem99,Wij99,Sob00a,Hom01}. Based on its strong
aperiodic variability, QPOs and X-ray spectrum, XTE~J1550-564 was
considered as a promising black hole candidate. Radio jets with
apparent superluminal velocities were observed after the strong
X-ray flare in 1998 September \citep{Han01}, indicating that
XTE~J1550-564 was another microquasar.

Subsequent optical observations showed that the dynamical mass of
the compact object is $10.5\pm1.0$ solar masses, indicating that the
compact object in XTE~J1550-564 is a black hole, its binary
companion was found to be a low-mass star and its distance was
estimated as about 5.3 kpc \citep{Oro02}. With {\sl Chandra}
Observatory, \citet{Cor02} discovered a large-scale,
relativistically moving and decelerating radio and X-ray emitting
jet from XTE~J1550-564.

After the major and complex outburst in 1998--1999, the source
experienced subsequent dim outbursts in 2000 \citep{Smi00,Mas00},
2001 \citep{Tom01,Jai01}, 2002 \citep{Swa02} and 2003 \citep{Dub03}.

X-ray temporal and spectral variabilities carry valuable information
about the emission and propagation process of high-energy photons
around the central black hole in a black hole X-ray binary system.
\citet{Neg94} constructed the averaged peak-aligned shot profile
with {\sl Ginga} observation data of the galactic black hole
 Cyg~X-1 in the hard state. With {\sl RXTE}/PCA data
and an improved shot detection algorithm, \citet{Fen99} studied the
evolution of hardness ratio of (13--60\,keV)/(2--6\,keV) in the
hard, soft, and transition states of Cyg~X-1. With the identical
method of obtaining the averaged peak-aligned shot profile,
\citet{Liu04} studied the spectral variability in the energy band
above $\sim 10$\,keV in different states of Cyg~X-1. To obtain
statistically significant results with the shot search technique in
time domain by \citet{Fen99}, sufficiently long observation and high
source flux are necessary in order to suppress noise, thus limiting
the application of the algorithm in other X-ray sources. Fortunately
XTE~J1550-564 was an exceptionally bright X-ray source in its 98/99
outburst and it was intensively observed a series of almost daily
pointed RXTE observations, which produce a rich dataset appropriate
for applying the above mentioned shot searching and analysis method.

In this work we make energy resolved temporal analysis of
XTE~J1550-564 with the {\sl RXTE}/PCA observation performed in
September 8, 1998, just after the source was discovered and in the
initial X-ray rising phase. We present temporal and spectral
property of averaged superposed shot, which has been studied in
Cyg~X-1 \citep{Neg94,Fen99,Neg01,Liu04} but almost no statistically
significant result has been obtained in other sources. Energy
resolved timing analysis of correlation is also made. We compare the
different spectral variabilities of XTE~J1550-564 and Cyg~X-1, and
discuss the possible physical interpretations related with such
phenomenon.

\section{SHOT PROPERTIES}

\subsection{Spectral State on the Rise to 98/99 Outburst}

The shot analysis has been progressed for different states of
Cyg~X-1. We find it is not easy to do the same with the data in
other sources, especially in their low/hard state (LS), due to the
lack of the observations with sufficiently high count rates, high
signal to noise ratio and long exposure which are capable of
producing statistically significant results. The rise to 98/99
outburst in XTE~J1550-564, however, gives an uncommon LS of
comparable brightness to Cyg~X-1, which is appropriate for shot
analysis and comparison with the results from Cyg~X-1 in LS.

The 98/99 outburst of XTE~J1550-564 exhibits a ''double-peaked ''
profile, separated by a minimum occurring around December 3, 1998.
\citet{Sob99} conclude that XTE~J1550-564 is observed in the very
high (VHS), high/soft (HS) and intermediate canonical outburst (IS)
states of black hole X-ray novae. Accurately, the first half of the
outburst from September 18 to October 27, 1998 has the spectra
dominated by the power-law component with large photon index, and is
identified as VHS. After October 27 the power-law component weakens
rapidly and the disk component begins to dominate, which is
consistent with HS. However, during this time the source
occasionally exhibits QPOs and those observations are considered as
IS.

In more detail, \citet{Wu02} divide the first peak of 98/99 outburst
into five phases, named after the corresponding stages in the {\sl
RXTE}/PCA X-ray light curve: (1) fast rising, (2) slow rising, (3)
flare, (4) postflare plateau and (5) X-ray decline. The first phase,
which lasts from September 7 to 8, is the onset of the outburst and
characterized by an impulsive rise in the hard X-rays (above
20\,keV) and an exponential-like rise in the soft X-ray (2-12\,keV
for {\sl RXTE}/PCA).  \citet{Wil01} study the spectral properties of
the first 14 observation of the rise to outburst phase from
September 7 to September 16. Using both the PCA and HEXTE instrument
aboard {\sl RXTE}, they find that the 3--200\,keV spectra smoothly
pass from standard LS to VHS without encountering classic HS.
Initial spectrum can be fitted adequately by a disk black body and a
thermal Comptonization model which includes reflection, and it is
very similar to the classic LS spectrum seen in many Galactic Black
Holes (little or no Black Body and a large Compton component). The
power spectrum of these data shown by \citet{Cui99} is also vary
similar to LS power spectra. With the further comparison with the
hardest LS spectrum of Cyg~X-1 observed by $RXTE$, they conclude
that in the first part of the rise, XTE~J1550-564 was in a classic
LS.

Therefore, the first two observations on the initial rise in
September 7 to 8, fulfill our necessary of LS with high flux. They
are similar in terms of their spectra \citep{Wil01} and power
spectra \citep{Cui99}. We pick the second one because of its longer
exposure time ($\sim 5000$ s) than the first ($\sim 1000$ s). So
that the data chosen by this work are just from the observation in
the next day of the detection of source while in its LS. The Obs. ID
of data is 30188-06-03-00, which is part of a guest observer program
with results reported by \citet{Cui99}.

\subsection{Shot and Spectral Variability}

 Light curves with 1\,ms time bin in different energy bands are
extracted from {\sl RXTE}/PCA Event mode (covering 13--60\,keV) and
Single-Bit mode (covering 2--13\,keV) of Obs. ID 30188-06-03-00,
with FTOOLS package. GTI criteria in the standard product by {\sl
RXTE} Guest Observer Facility (GOF) is applied, which means the data
are selected for analysis when the source is observed at the
elevation angle larger than $10\,^{\circ}$, the offset pointing is
less than $0.02\,^{\circ}$, at least one PCU is on, and after thirty
minutes since the peak of SAA (South Atlantic Anomaly) passage. The
98/99 outburst of XTE~J1550-564 is bright, and the averaged
background contribution for high energy band ($>10\sim 20$\,keV) is
at 10\% lever and less than 1\% for low energy band ($<10\sim 20$\,keV).
Therefore the background is negligible for the study of averaged
shot features. The effect of dead time is also neglected since the
count rate is not very high and the time bin is much larger than
{\sl RXTE} dead time of $10\,\mu s$.

\begin{figure}
\epsscale{0.9} \plotone{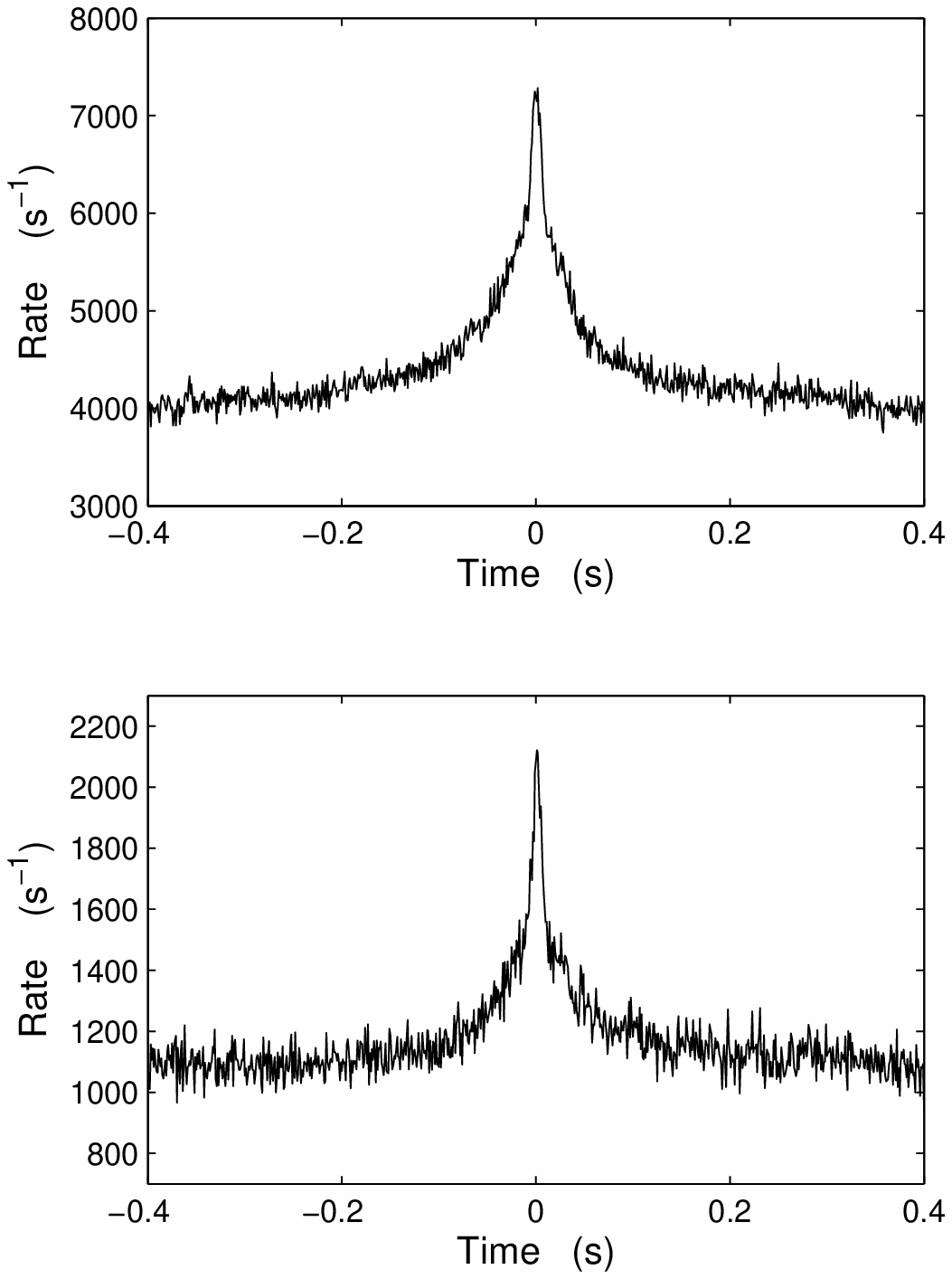} \caption{Average shot profile of
XTE~J1550-564 in LS ($RXTE$/PCA Obs. ID 30188-06-03-00. Start time
1998-09-08 00:09:39, MJD 51064.0067013888911; stop time 1998-09-08
02:36:07, MJD 51064.1084143518528). {\it Top panel}: 2--13\,keV.
{\it Bottom panel}: 13--60\,keV.\label{fig1}}
\end{figure}

Shots are searched from 1\,ms time bin light curves by using the
algorithm proposed by \citet{Fen99}. A light curve with 1\,ms time
bin is first merged into a larger bin of 10\,ms. Then the bin with
count $C_p$ larger than the neighboring bins is selected as a peak
candidate. In the neighboring 1\,s on both sides of each candidate,
we search for the bins with counts $C_1$ and $C_2$ so that the
condition $ C_p > C_{1, 2} + 2\sqrt{C_p+C_{1, 2}}$ is satisfied. If
and only if the number of bins with $C_1$ and $C_2$ is larger than a
certain criterion, the candidate peak is selected as a potential
shot peak. Then each potential shot peak and its neighboring bins on
both sides are divided into 30 bins with a time bin of 1 ms. A shot
is finally selected by the criterion that its count should be the
maximum within the 30 ms and 2 times larger than the mean count of
the observation. The above process is performed in  energy bands of
2--13\,keV and 13--60\,keV respectively. A shot peak is finally
identified if it coincides in both energy bands within 30 ms. Many
shot peaks determined this way are aligned to obtain the averaged
shot profiles separately in each energy band, as shown in
Figure~\ref{fig1}.

\begin{figure}
\epsscale{1} \plotone{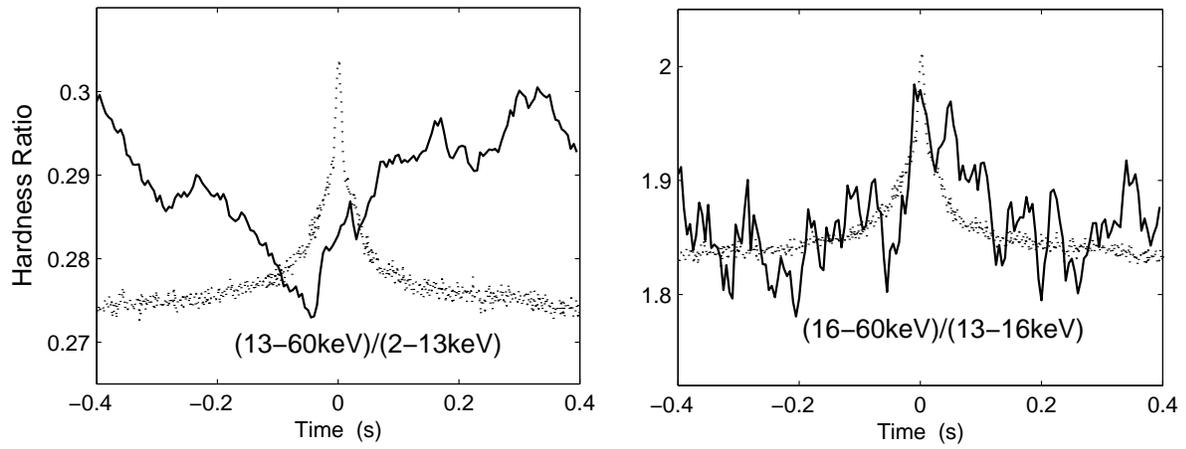} \caption{Flux and hardness ratio
profiles of average shot of XTE~J1550-564 in LS (Obs. ID
30188-06-03-00). {\it Dotted lines} are normalized shot flux
profiles; {\it solid lines} are hardness ratios. {\it Left panel} :
(13--60\,keV)/(2--13\,keV). {\it Right panel} :
(16--60\,keV)/(13--16\,keV). \label{fig2}}
\end{figure}

In order to study the spectral variability during an averaged shot,
for each time bin within a 400 ms interval around the shot peak, the
hardness radio is calculated with
\begin{displaymath}h=\frac{\sum_{i=1}^N f_A(i)/f_B(i)}{N}\end{displaymath}
where $f_A(i),f_B(i)$ are the count rates for the $i$th selected
shot in energy band A and B respectively, and $N$ is the total
number of selected shots in the above process. Figure~\ref{fig2}
shows the total profiles of averaged shot in 2--60\,keV and hardness
ratio between different bands during the averaged shot.

We can see from Figure~\ref{fig2} that during a shot the hardness
ratio with respect to a soft band below $\sim 10$\,keV is negatively
correlated to the flux, namely the hardness ratio decreases as the
flux of shot increases and reaches minimum at the shot peak bin,
then rises with flux dropping. The variation of hardness ratio of
(16--60\,keV)/(13--16\,keV) during a shot is somewhat complex,
however, in the main part of shot, it shows a rise and reaches peak
when the flux of the shot peaks. Thus there is a possibility that
the hardness with soft band above $\sim 10$\,keV is positively
related to the light curve, which will be verified by the following
correlation analysis.

\section{CORRELATION BETWEEN HARDNESS AND INTENSITY}
\citet{Li99} proposed an algorithm to calculate the correlation
coefficient $r(h,f)$ between the hardness ratio $h$ and the total
intensity $f$ on a given timescale $\Delta t$. The light curve in
the studied observation period is divided into $N$ segments with
duration of $10\Delta t$ each. The correlation coefficient
$r_{k}(i)$ of the $k$th segment is calculated with the equation
\begin{displaymath}
 r_{k}(h,f) =\frac{\displaystyle\sum_{i=1}^{10}{(h(i)-\bar h)(f(i)-\bar
f)}}{\sqrt{\displaystyle\sum_{i=1}^{10}{(h(i)-\bar
h)^2}\displaystyle\sum_{i=1}^{10}{(f(i)-\bar f)^2}}}
\end{displaymath} and then their average $\bar r=\sum_{k=1}^N
r_{k}/N$ and standard deviation $\sigma(\bar r)=\sqrt{\sum_{k=1}^N
(r_{k}-\bar r)^2/(N-1)}$.

\begin{figure}
\epsscale{1}\plotone{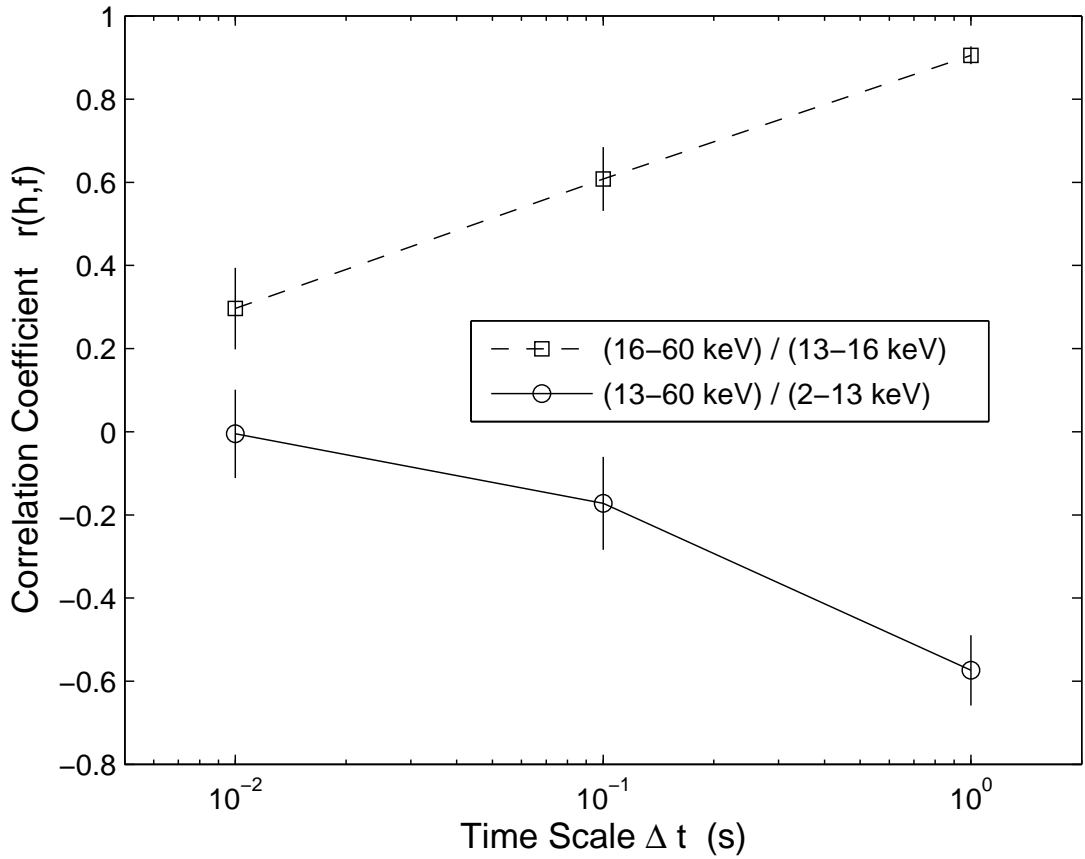} \caption{Correlation coefficient
between hardness ratio and intensity vs. time scale of XTE~J1550-564
in LS (Obs. ID 30188-06-03-00). {\it Solid line} and {\it circle
mark} for hardness ratio of (13--60\,keV)/(2--13\,keV); {\it Dashed
line} and {\it square mark} for hardness ratio of
(16--60\,keV)/(13--16\,keV). \label{fig3}}
\end{figure}

The correlation coefficients between the intensity and hardness
ratio of (13--60\,keV)/(2--13\,keV) and (16--60\,keV)/(13--16\,keV)
are obtained on timescales of 0.01 s, 0.1 s and 1 s respectively.
The results are shown in Figure~\ref{fig3}. From Figure~\ref{fig3}
the difference of coefficients with regard to the soft bands below
and above $\sim 10$\,keV is clear: for the former the coefficients
 are negative and decrease monotonically with timescale increasing,
and for the latter are positive and increase monotonically.

\section{COMPARISON WITH CYG~X-1}
 We summarize the main features in X-ray spectral variability in
XTE~J1550-564 revealed by our time domain analysis as follows: (1)
During an averaged shot, the hardness ratio with regard to a soft
band below $\sim 10$\,keV has a ''V'' shape, i.e. shots are softer
than the steady emission and the hardness ratio in the hard tail of
the energy spectrum above $\sim 10$\,keV has complicated evolution
and peaks when the flux peaks. (2) The correlation coefficients
between the intensity and hardness ratio with regard to the soft
band below $\sim 10$\,keV on timescales between 0.01 and 1 s are
negative and decreases monotonically as timescale increases; in
contrast, the correlation in the energy band above $\sim 10$\,keV is
positive and increases with timescale increasing. The correlation
analysis is made for the temporal variability of the whole light
curve instead of only shot, and its results in Figure~\ref{fig3} are
qualitatively consistent with the results from the shot analysis.
The monotonic increase or decrease with increasing time scale may
indicate the effect of uncorrelated noise being weakened on larger
timescales.

The shot analysis technique has been applied to Cyg~X-1 in different
spectral states and energy bands with $RXTE$/PCA data
\citep{Fen99,Liu04}. That the average duration of shots of
XTE~J1550-564 in the high energy band is shorter than what observed
in the low energy band, shown in Figure~1, is consistent with the
result in Cyg~X-1. We notice from Figure~\ref{fig2} that there
exists a pre-peak drop of hardness with soft band above $\sim
10$\,keV during a shot, which has also been observed in the
transition and low states in Cyg~X-1, and explained by the more
effective cooling in the disturbance.

\begin{figure}
\epsscale{1}\plotone{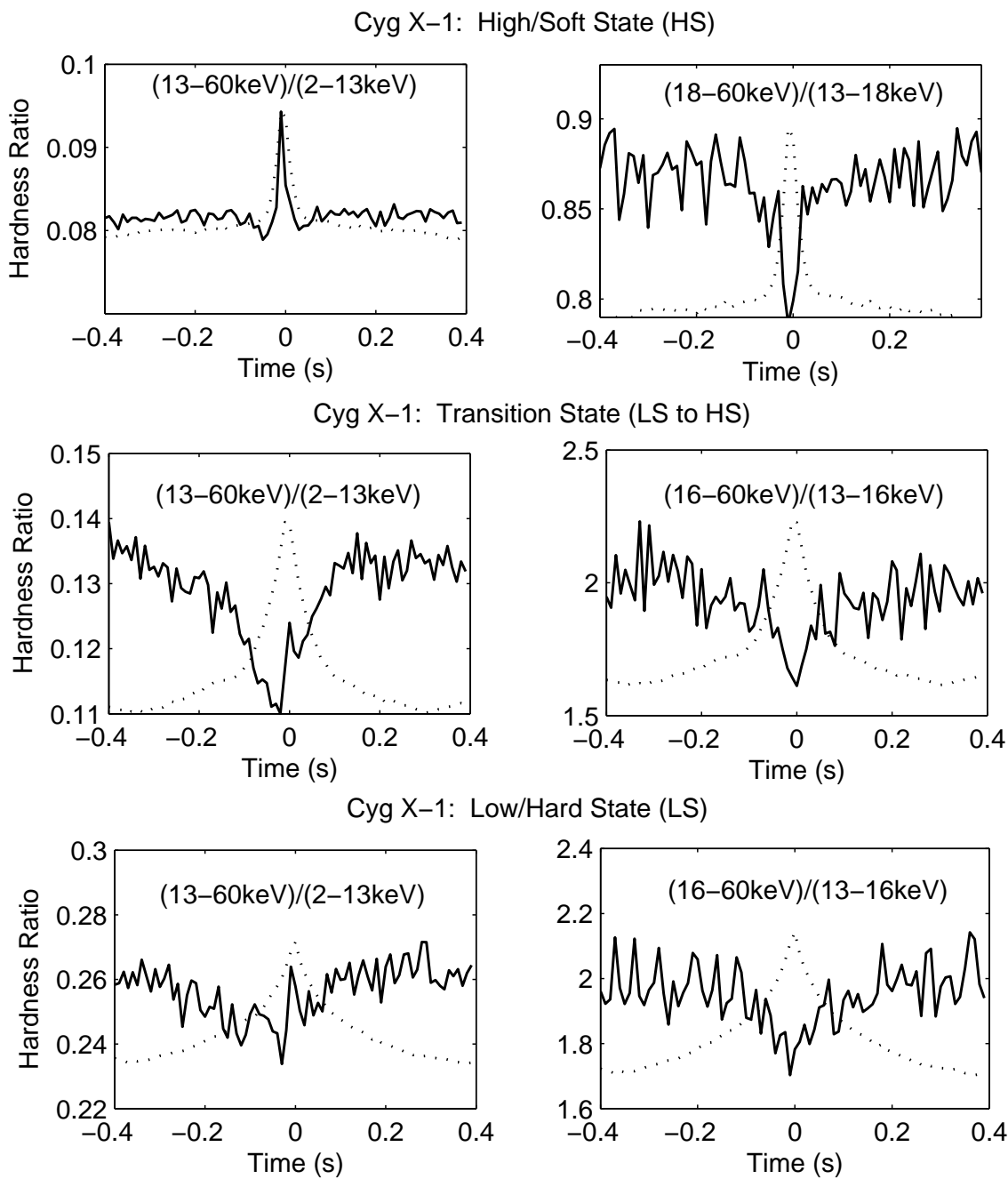} \caption{ Flux and hardness ratio
profiles of average shot from {\sl RXTE} data of Cyg~X-1 in
different states (adapted from \citet{Liu04}).  {\it Dotted line}:
normalized shot flux profile;  {\it Solid line}: hardness ratio.
 {\it Top row}: high/soft state, Obs. ID 10512-01-08-00;  {\it Middle
row}: hard to soft transition, Obs. ID 10412-01-01-00,
10412-01-03-00; {\it Bottom row}: low/hard state, Obs. ID
10236-01-01-03, 10236-01-01-04.   {\it Left column}:
(13--60\,keV)/(2--13\,keV); {\it Right column}: (16--60
keV\,)/(13--16\,keV).  \label{fig4}}
\end{figure}

\begin{figure}
\epsscale{1}\plotone{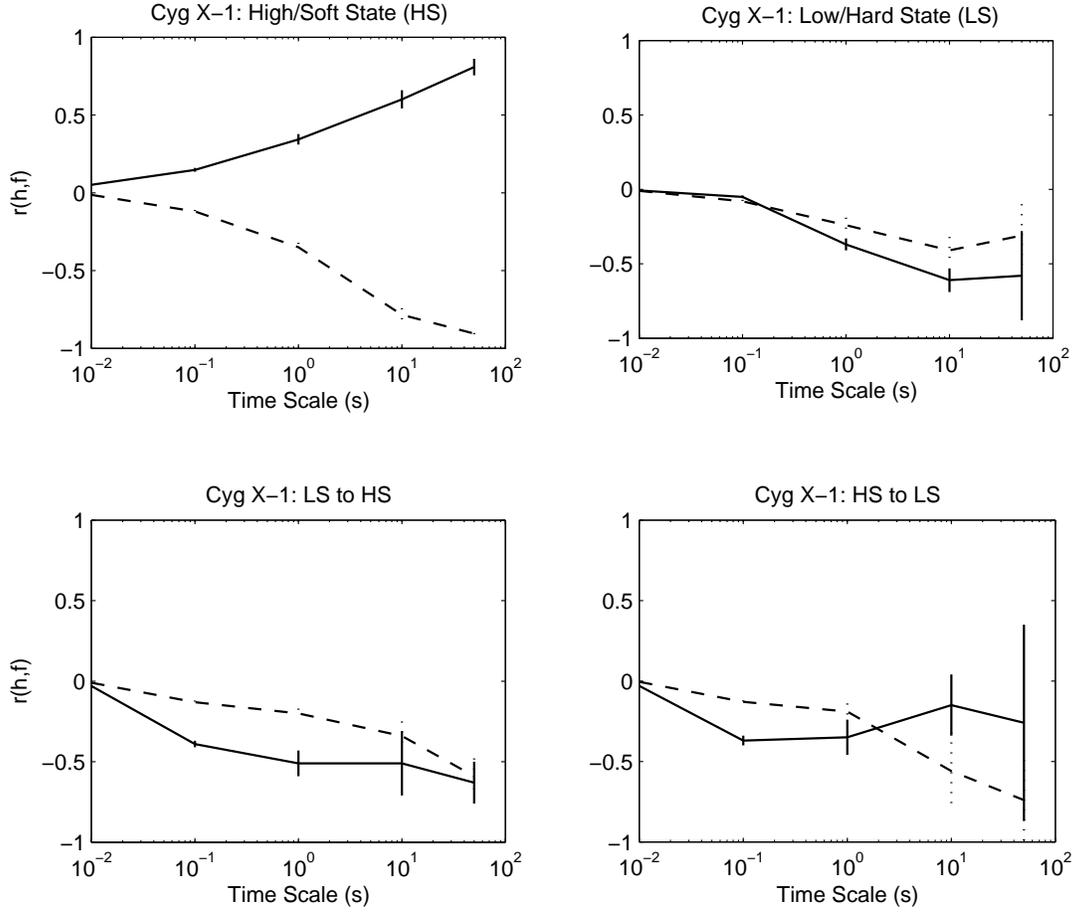} \caption{ Correlation coefficients
between hardness and intensity vs. time scale in Cyg~X-1 (adapted
from \citet{Liu04}).  {\it Solid line}: hardness in
(13--60\,keV)/(2--5\,keV);  {\it Dashed line}: hardness in
(16--60\,keV)/(13--16\,keV).  {\it Upper-left panel}: high/soft
state (HS); {\it Upper-right panel}: low/hard state (LS);  {\it
Lower-left panel}: transition state (LS to HS);  {\it Lower-right
panel}: transition state (HS to LS). \label{fig5}}
\end{figure}

\begin{figure}
\epsscale{1}\plotone{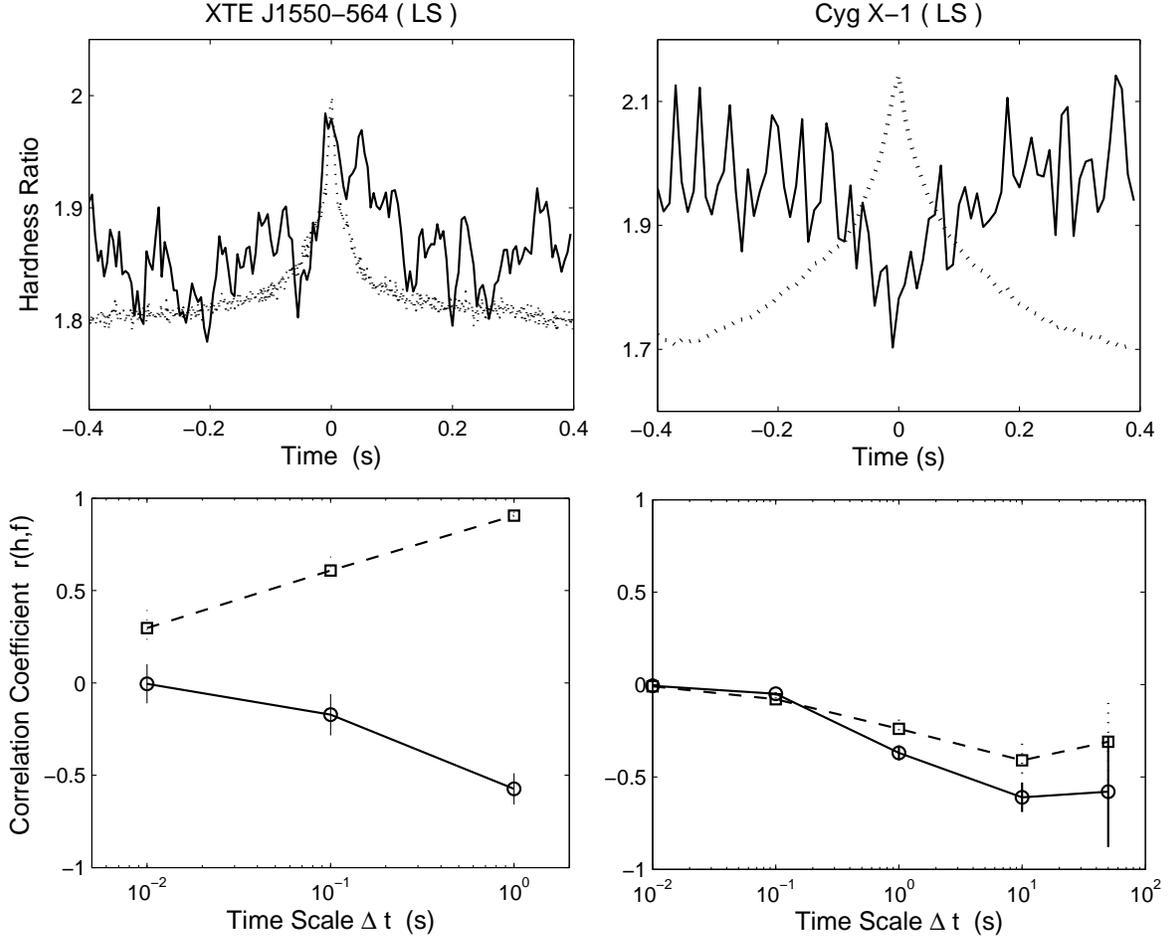} \caption{ Comparison of both sources in
LS. {\it Left column}: XTE~J1550-564; {\it Right Column}: Cyg~X-1.
{\it Top row}: flux and hardness ratio (16--60\,keV)/(13--16\,keV)
profiles of average shot; {\it Bottom row}: correlation coefficients
between hardness and intensity vs. time scale. {\it Solid line} and
{\it circle mark} correspond to hardness ratio of
(13--60\,keV)/(2--13\,keV);  {\it Dashed line} and {\it square mark}
correspond to hardness ratio of (16--60\,keV)/(13--16\,keV).\label{fig6}}
\end{figure}

Figure~\ref{fig4} and Figure~\ref{fig5} are adapted from
\citet{Liu04} showing hardness profiles of averaged shot and
timescale distribution of correlation coefficient between hardness
and intensity for Cyg~X-1 in different spectral states. With regard
to the soft band of below $\sim 10$\,keV, the spectral evolution
during an averaged shot of XTE~J1550-564 in LS (the left panel in
Figure~\ref{fig2}) is similar with that of Cyg~X-1 in LS (the
bottom-left panel in Figure~\ref{fig4}): the average hardness ratio
(13-60\,keV/2-13\,keV) during a shot is in general lower than that
of the steady component and a sharp rise in the average profile of
hardness ratio is appeared at about the peak of shot flux. The
feature of timescale distribution of correlation coefficient between
X-ray hardness and intensity for XTE~J1550-564 (the solid line in
Figure~\ref{fig3}) is also similar with that for Cyg~X-1 (the solid
line in the upper-right panel of Figure~\ref{fig5}): the
correlations between the hardness ratio $h$ and intensity $f$ on
time scales between 0.01\,s and 1\,s are negative and the
correlation coefficients $r(h,f)$ decrease monotonically along with
timescale increasing. In contrast, for the hard band above $\sim
10$\,keV, the shot spectral evolution and timescale distribution of
hardness-intensity correlation coefficient are completely different
between the two sources when they both in LS, as shown in
Figure~\ref{fig6}: when flux peaks during a shot, the average
hardness profile bottoms out for Cyg~X-1 (see the upper-right panel
of Figure~\ref{fig6}) but peaks for XTE~J1550-564 (see the
upper-left panel of Figure~\ref{fig6}); the signs and timescale
dependence of hardness-intensity correlation coefficient in the hard
band above $\sim 10$\,keV are similar with that of
hardness-intensity correlation coefficient with regard to the soft
band below $\sim 10$\,keV for Cyg~X-1 (see the bottom-right panel of
Figure~\ref{fig6}), but completely different for XTE~J1550-563 (see
the bottom-left panel of Figure~\ref{fig6}).

\section{DISCUSSION}
 It is natural to assume that X-ray shots are most probably
produced at the innermost region of the cold disk where it joins the
hot corona, which is a most turbulent region, and the steady
component around a shot is a global average of emission from
different regions of the hot corona \citep{Li99}. The evolution of
hardness ratio with regard to a soft band below $\sim 10$\,keV
should depend on the comparison between the local temperature where
shot is produced and the global average temperature of the corona.
For LS the optically thick disk is truncated at a larger distance
and joins to a spherical corona around the black hole \citep{Esi97}.
Shots should be softer than the steady emission since they emerge
and are Comptonized at the outer part of the corona with temperature
lower than the average of the total corona. Such interpretation
coincides with the observations for both Cyg~X-1 and XTE~J1550-564.

For the hard band above $\sim 10$\,keV, two different processes may
influence the spectral evolution during a shot, i.e. Comptonization
of soft photons by hot electrons in the corona and reflection of
hard photons by the accretion disk or other reflector. The hard
power-law like spectrum of a black hole binary is produced by
thermal Comptonization: hard emission above $\sim 10$\,keV is
produced by inverse Compton scattering of soft seed photons by hot
electrons of temperature $kT_e\le 100$\,keV \citep{Sun79}. Besides
the Comptonization, irradiation of disk and other dense material by
the hard X-rays gives rise to a broad hump-like spectrum around
20--30\,keV (the Compton reflection hump) \citep{Geo91,Rey96}.

When a seed photon of 10\,keV collides with a relativistic electron
in the hot corona with energy 100\,keV, the energy of the seed
photon will be lifted to $\sim 14$\,keV. Most scattered photons
escaping from the hot corona undergo only one collision. If thermal
Comptonization is the dominant process to produce the shot energy
spectrum above $\sim 10$\,keV, the hardness ratio in respect to the
13--16\,keV band will decrease when the number of seed and scattered
photons increases, and an anti-correlation between the hardness
ratio (16--60\,keV)/(13--16\,keV) and the hard band intensity during
a shot will be expected as observed in Cyg~X-1 \citep{Liu04}.

\begin{figure}
\epsscale{0.7}\plotone{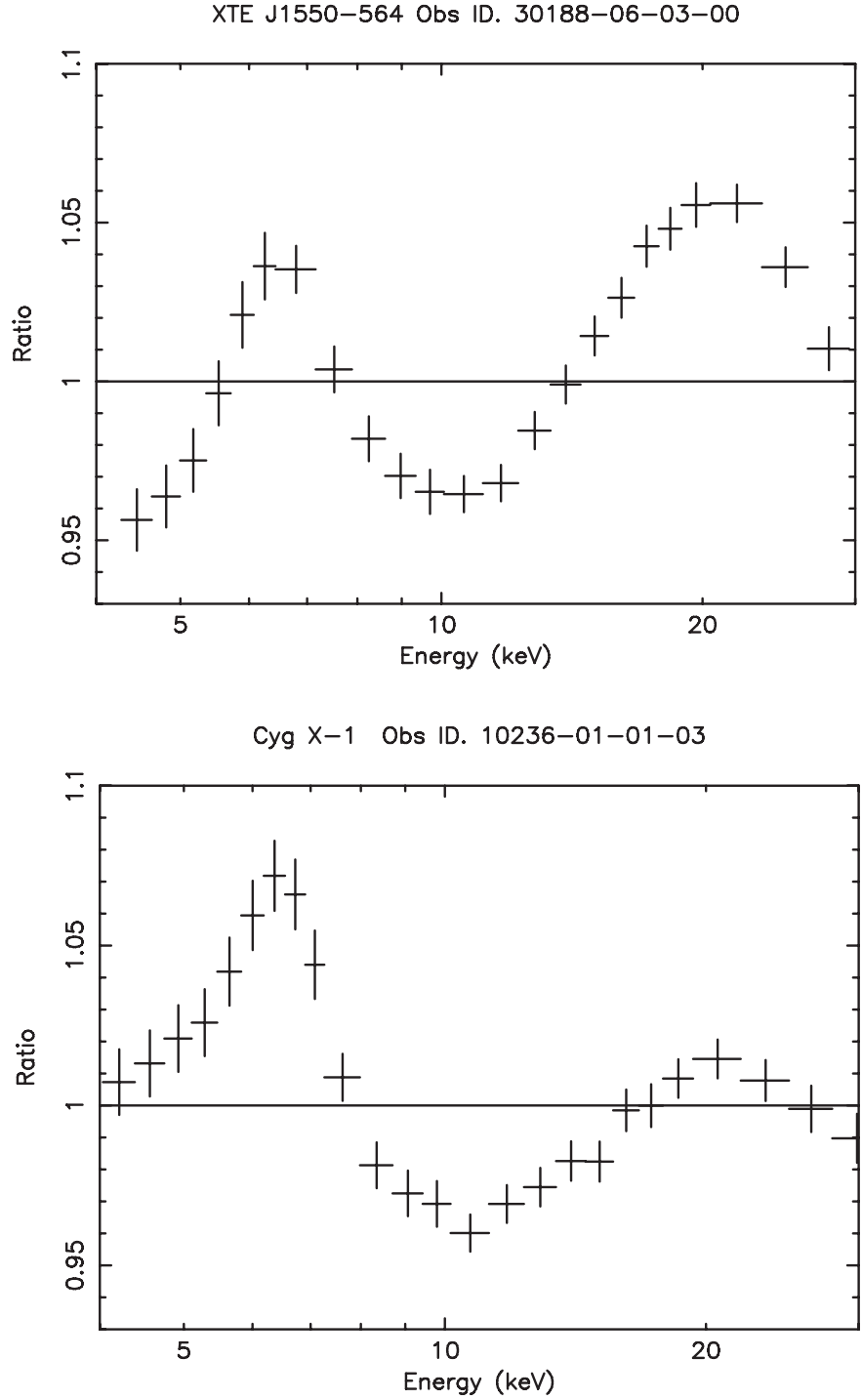} \caption{ Ratio of data to continuum
when a simple power law is fitted to energy spectra. {\it Upper
panel}: XTE~J1550-564, Obs. ID 10512-01-08-00; {\it Lower panel}:
Cyg~X-1, Obs. ID 10236-01-01-03. \label{fig7}}
\end{figure}

\begin{figure}
\epsscale{0.9}\plotone{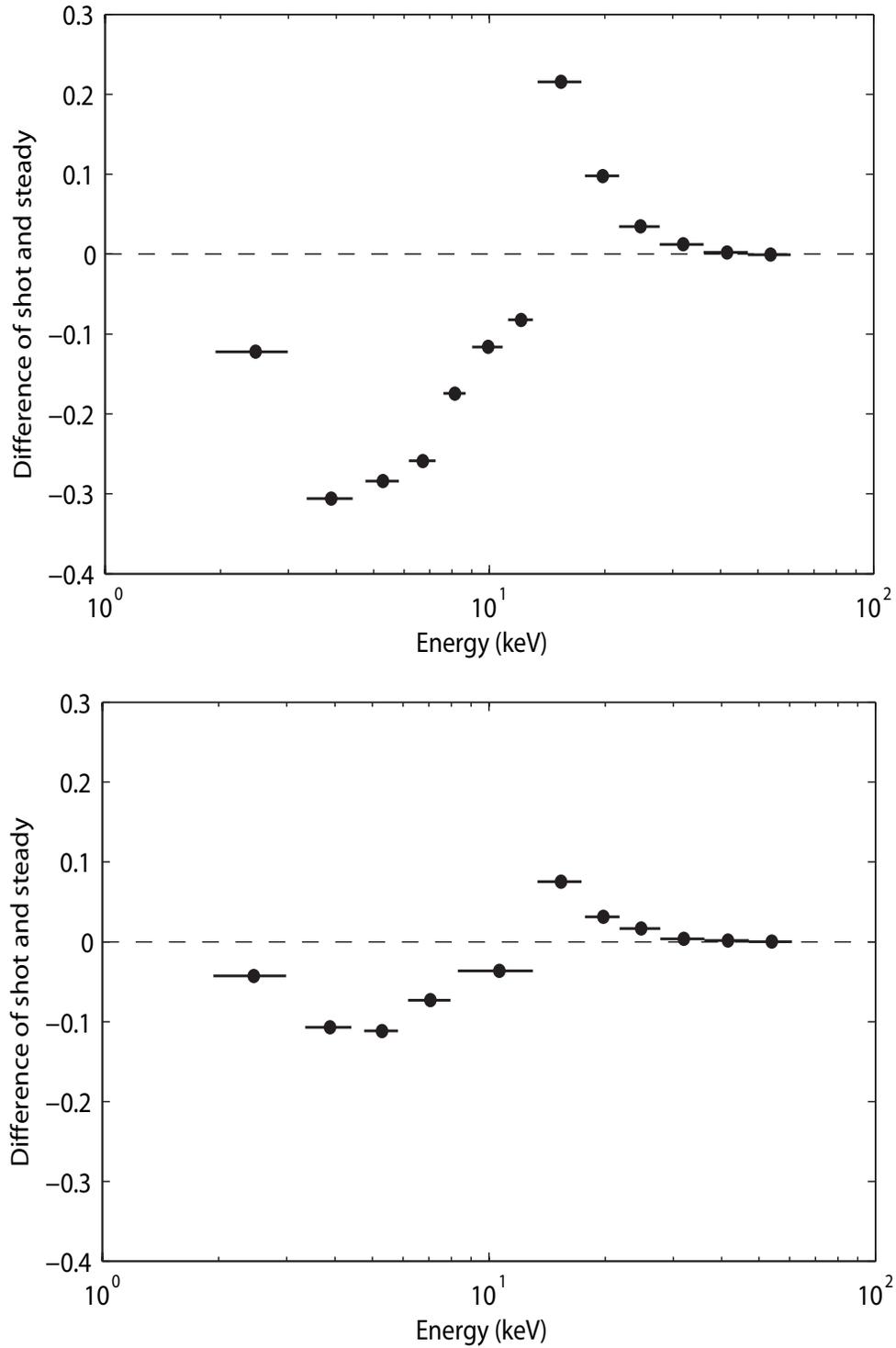} \caption{ Difference spectrum
obtained by subtracting the normalized counts spectrum of steady
component from that of shot component.{\it Upper panel}:
XTE~J1550-564, Obs. ID 10512-01-08-00; {\it Lower panel}: Cyg~X-1,
Obs. ID 10236-01-01-03. \label{fig8}}
\end{figure}

On the other hand, the reflection hump around 20--30\,keV should
increase the hardness ratio (16--60\,keV/(13--16\,keV). The observed
behavior of hardness (16--60\,keV)/(13--16\,keV) during shot in
XTE~J1550-564 may indicate that reflection component is an important
part in the energy band above $\sim 10$\,keV of observed shot
photons. To check this scenario, firstly we inspect the energy
spectrum of XTE~J1550-564 and compare with that of Cyg~X-1.
Figure~\ref{fig7} shows the ratio of data to continuum when a simple
power law is used to fit spectra of XTE~J1550-564 and Cyg~X-1
respectively. From Figure~\ref{fig7} one can see that the reflection
component around $\sim 20$\,keV seems more prominent in
XTE~J1550-564 than in Cyg~X-1, which is similar to what has been
seen from X-ray observations of active galactic nuclei (AGN), e.~g.
the bright Seyfert 1 galaxy MCG-6-30-15 \citep{Lee98,Bal03}, and
that might attribute to the different hardness evolution behaviors
between XTE~J1550-564 and Cyg~X-1. Secondly, we investigate the
spectrum of shot component picked out by the method described in
section 2, as well as steady component which is the residual of the
light curve after getting rid of selected shots. As the
characteristic timescale of shots is $\sim 0.1$\,s, the standard 2
mode with time step of 16 s is not appropriate to extract the shot
spectra. We use the data of Generic Binned and Event mode with
higher time resolution to obtain spectra at the cost of reducing
energy resolution. The counts spectra of both shot and steady
components are normalized by the total observed counts in a certain
energy bin. The difference spectra of XTE~J1550-564 and Cyg~X-1
obtained by subtracting the normalized counts spectrum of steady
component from that of shot component are shown in
Figure~\ref{fig8}. From Figure~\ref{fig8} we can see that for both
sources the proportion of Fe K$\alpha$ line at $\sim 6.4$\,keV in
shot component is less than that in steady component, and in the
hard band more prominent reflection photons around $\sim 20$\,keV
exist in shots than in steady emission. It seems from Figures 7 and
8 that for XTE~J1550-564 the more fraction of reflection component
around $\sim 20$\,keV may possibly interpret the observed hardness
evolution profiles in the energy band above $\sim 10$\,keV during
the averaged shot.

It is not a surprise to see the similarity in the presence of a
strong reflection continuum above 10\,keV between the microquasar
XTE~J1550-564 (Figure 7) and AGNs, e.~g. MCG-6-30-15 (Figure 4 in
\citet{Lee98}; Figure 2 in \citet{Bal03}). But to interpret the
positive correlation between intensity and hardness ratio in the
hard band above $\sim 10$\,keV by reflection effect during an
average shot and the negative correlation in regard to the soft band
below $\sim 10$\,keV by the local temperature where shot is produced
being lower than the average of the corona, we need further assume
that the strong reflection continuum is a characteristic of  X-ray
shots and the fluorescent Fe line is mainly from the steady
component, which is indeed indicated by Figure 8. A possible model
to meet the above requirement is schematically represented by Figure
9. In this model, the Fe line is produced from steady X-rays in the
hot corona irradiating the accretion disk and the Compton hump above
10\,keV from reflection of X-ray shots by reflector far away from
the corona where the steady emission is weak and jet-like shots may
still intense enough. The possible electro-magnetic origin of X-ray
shots, e.~g. magnetic flares \citep{Bel99,PF99} or plasma columns
produced by pinch discharge \citep{Alf81,Wu05} may produce
collimated emission to some extent. With this model, the observed
difference of hardness evolution in an average shot between the two
sources comes from that there exist outer reflectors for the
microquasar XTE~J1550-564 (at least in the initial X-ray rising
phase) which is more than Cyg~X-1. Further inspecting spectral
evolution, polarization and other properties for average and even
individual shot of GBHC and AGN with next generation X-ray explorers
is necessary for understanding the process of shot production and
the environment around the central black hole.

 \acknowledgments The authors thank Prof. S. N. Zhang and the referee for helpful
comments and suggestions. This work is supported by the National
Natural Science Foundation of China. The data analyzed in this work
are obtained through the HEASARC on-line service provided by the
NASA/GSFC.

\clearpage

\clearpage

\clearpage

\clearpage

\clearpage

\clearpage

\clearpage

\clearpage

\clearpage
\begin{figure}
\epsscale{1}\plotone{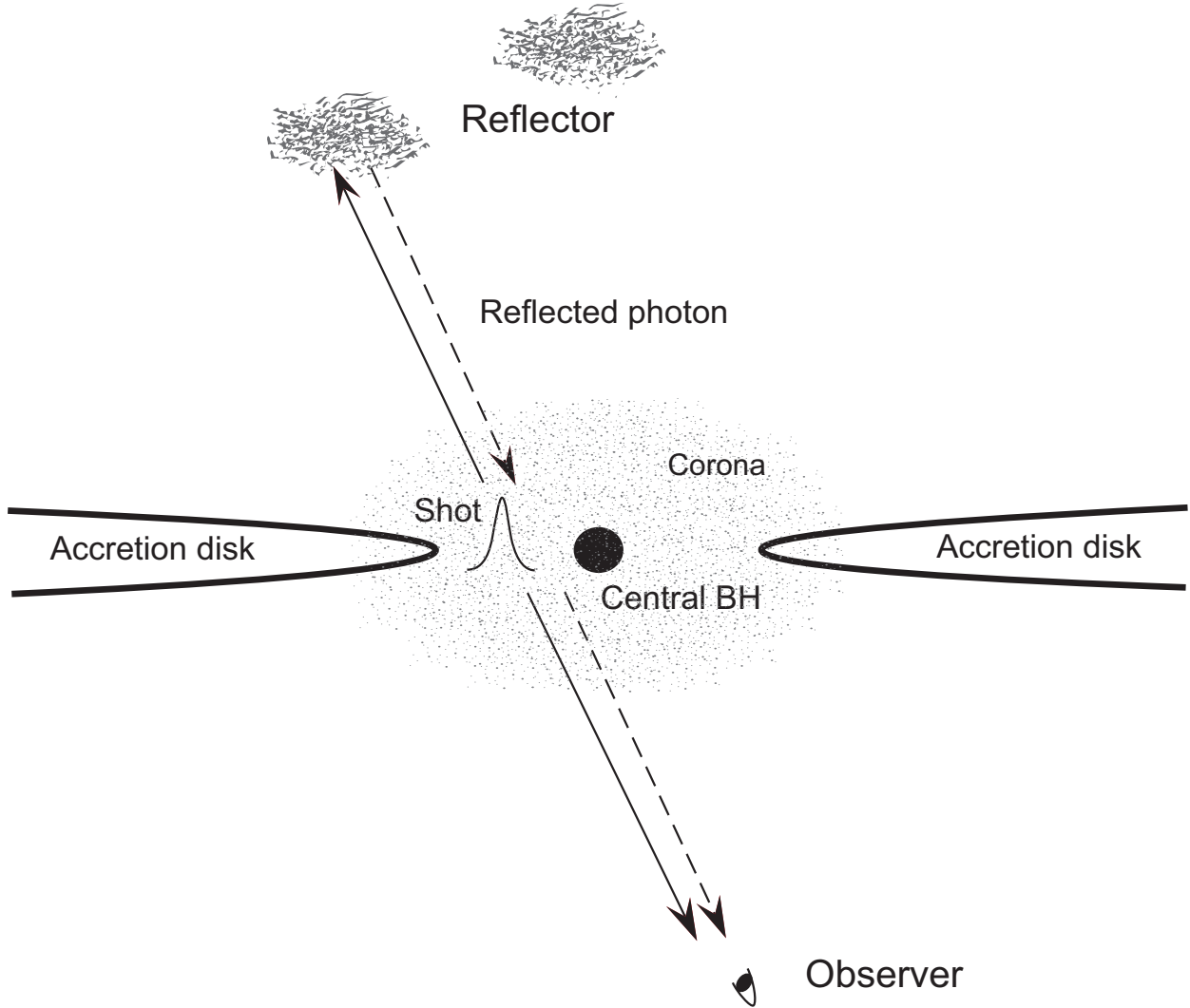} \caption{ A schematic representation
of a possible model to interpret the spectral evolution in X-ray
shots of XTE~J1550-564.  Shot is produced by electromagnetic
process and ejects collimated light beams in two opposite
directions, shown as the {\it solid lines with arrows} in figure.
There exist clumps of matters around central BH which reflect a
fraction of collimated light, shown as {\it dashed lines with
arrows}. What we detect are photons directly from the beam
approaching us as well as the reflected photons from receding one.
The direction of beams are random and only those parallel to the
sight line can be observed. \label{fig9}}
\end{figure}

\end{document}